# On derivation of Wigner distribution function

S Khademi

Department of Physics, 5th Km of Tabriz Road, Zanjan University, Zanjan, Iran

Email:skhademi@mail.znu.ac.ir  and  siamakkhademi@yahoo.com

**Abstract**. Wigner distribution function has much importance in quantum statistical mechanics. It finds applications in various disciplines of physics including condense matter, quantum optics, to name but a few. Wigner distribution function is introduced by E. Wigner in 1932. However, there is no analytical derivation of Wigner distribution function in the literatures, to date. In this paper, a simple analytical derivation of Wigner distribution function is presented. Our derivation is based on two assumptions, these are A) by taking the integral of Wigner distribution function, with respect to configuration space, the momentum space distribution function is obtained B) WDF is real. Similarly, and in addition to Wigner distribution function, the distribution function of Sobouti-Nasiri, which is imaginary, is also derived.

## 1. Introduction

In 1932 Wigner introduced a distribution function (DF), with an excellent penetration [1, 2]. He also derived what is known as Wigner equation, which governs the dynamics of Wigner distribution function (WDF) [1-3]. This DF, inherently, encompasses all requirements of both quantum and statistical mechanics, simultaneously. In this context, the basis of quantum statistical mechanic formulation is constructed [1-3]. The main problem with the Wigner formulation is that WDF turns out to be negative [4]. Later, many authors proposed some imaginary DFs, which found a wide use in various field of physics [5-9]. They have argued the DFs not only can have negative parts, but also may essentially be imaginary functions [9]. The well-known (imaginary) DFs are sometimes obtained directly from WDF.

Furthermore, Sobouti and Nasiri obtained an imaginary DF by extension of phase s pace [10, 11]. The Sobouti-Nasiri DF is derived from the basic physical principles. Extension of Lagrangian and Hamiltonian methods as well as canonical quantization is used, to obtain extended phase space (EPS) formulation. They have shown that the EPS formulation is equivalent to Wigner approach [10-13]. They obtain WDF, by applying a unitary operator to Sobouti-Nasiri DF. Their method may be interpreted as a derivation of WDF from basic physical principles, in a complicated manner.

The aim of this paper is to derive quantum statistical DFs using a method simpler than EPS formulation. For this, one may consider two basic assumptions. At first one may assume that, *the distribution function in momentum space is obtained by integration of the phase space distribution functions* and seconds *WDF can only be real.* The first assumption is holds for all DFs, while the second one is just for WDF.

In this paper after an introduction in section one a brief review of Wigner formulation is given in section 2. The next section is devoted to derivation of WDF. Finally, the concluding remarks are presented in the last section.

## 2. A Brief Review of Wigner formulation

Wigner distribution function is introduced by

$$P_W(p,q) = \frac{1}{2\pi\hbar} \int dy \langle q - y/2 | \hat{\rho} | q + y/2 \rangle \exp(\frac{ipy}{\hbar}), \qquad (1)$$

in Dirac's notation [1], where $\hat{\rho}$ is density operator. Similarly, in terms of quantum wave function $\psi$ (which is derived from Schrödinger equation), WDF is given by [2]

On derivation of Wigner distribution function

$$P_W(p,q) = \frac{1}{2\pi\hbar} \int dy \, \psi^*(q+y/2)\psi(q-y/2)\exp(\frac{ipy}{\hbar}). \quad (2)$$

An averaging rule is used to obtain the expectation value of the physical quantities, say $\hat{A}$, as

$$<\hat{A}> = \int dpdq P_W(p,q) A(p,q), \quad (3)$$

The physical quantities $A(p,q)$ are obtained from the Weyle-Wigner transformation [2]

$$A(q,p) = \int \langle q - \tfrac{1}{2}y |\hat{A}| q + \tfrac{1}{2}y \rangle e^{\frac{ipy}{\hbar}} dy, \quad (4)$$

where $\hat{A}$ is, in general, an ordinary quantum mechanical operator corresponding to the physical quantities. All DFs, such as WDF, should be normalized

$$\int dpdq P_W(p,q) = 1. \quad (5)$$

It is easy to see that Wigner DF is governed by Wigner equation [1, 2]

$$i\hbar \frac{\partial P_W}{\partial t} = -\frac{p}{m}\frac{\partial P_W}{\partial q} + \sum_{n=0} \frac{1}{(2n+1)!}(\frac{\hbar}{2i})^{2n} \frac{\partial^{2n+1}V}{\partial q^{2n+1}} \frac{\partial^{2n+1}P_W}{\partial p^{2n+1}}, \quad (6)$$

where $V$ is an scalar potential.

### 3. Derivation of WDF

Wigner introduced his distribution function, in 1932, but didn't present a derivation of WDF. Educationally, derivation of WDF is important, in understanding of quantum statistical mechanics.

Researchers are interested in obtaining the quantum DFs from a fundamental basis. To this end purpose, Sobouti and Nasiri present a method to formulate quantum statistical mechanics [10]. They investigated extension of phase space in Lagrangian, Hamiltonian and canonical quantization [10, 11] to construct the extended phase space (EPS) formulation corresponding to quantum statistical mechanics, in the same manner as that of Wigner. In the EPS formulation, an imaginary DF, $P_{SN}$ is obtained and WDF is derived by applying a unitary operator to Sobouti-Nasiri DF,

$$P_W(p,q,t) = e^{(\frac{1}{2})\frac{i\hbar\partial^2}{\partial p \partial q}} P_{SN}(p,q,t). \quad (7)$$

Here the subscript SN denotes instead of Sobouti-Nasiri DF. In EPS formulation the imaginary DF is derived as

$$P_{SN}(q,p) = \frac{1}{2\pi\hbar} \int <q|\hat{\rho}|q+y> e^{\frac{ipy}{\hbar}} dy. \quad (8)$$

Interested readers should see ref. 10.

Equation (7) is, in fact, a derivation of WDF in EPS formulation. Unfortunately, EPS derivation of WDF has some complication. To have a simpler derivation of WDF, two simple assumptions are considered:

*1) Distribution function in momentum space is obtained by integrating the distribution function*

$$P_{mom}(p) = \int dq P(p,q) = Tr(\hat{\rho}\delta(p-\hat{p})). \quad (9)$$

*2) WDF is real.*

The first assumption is used for all DFs, while the second one is considered just for WDF. But, the reality is not a necessary condition for DFs of quantum statistical mechanics. Many authors are introduced different DFs which have all expected properties of DFs [2].

On derivation of Wigner distribution function

To calculate the expectation of an arbitrary physical quantity $\hat{A}$, one may take the average of $A$ in phase space, using any DFs

$$<\hat{A}> = \int P(p,q) A \, dp \, dq. \tag{10}$$

One may also investigate the averaging rule in momentum space

$$<\hat{A}> = \int P_{mom}(p) A \, dp = Tr(\hat{\rho}\hat{A}), \tag{11}$$

where the correct DF in momentum space is defined by [2]

$$P_{mom} = Tr(\hat{\rho}\delta(p - \hat{p})). \tag{12}$$

To prove this definition let's substitute the definition of momentum space (12) in Eq. (11),

$$<\hat{A}> = \int P_{mom}(p) A \, dp = \int Tr(\hat{\rho}\delta(p - \hat{p})) A \, dp,$$
$$= \int dp \, dq <q|\hat{\rho}\delta(p - \hat{p})|q> A = \int dp \, dq \, dp' <q|\hat{\rho}\delta(p - \hat{p})|p'><p'|q> A,$$
$$= \int dq \, dp \, dp' <q|\hat{\rho}|p'><p'|q> A \delta(p - p') = \int dq <q|\hat{\rho}\hat{A}|q>,$$
$$= Tr(\hat{\rho}\hat{A}).$$

On the other hand, second simple assumption dictates

$$P_{mom} = \int dq \, P(p,q). \tag{13}$$

Comparing Eq. (13) and (12) gives

$$P_{mom}(p) = \int P(p,q) dq = Tr(\hat{\rho}\delta(p - \hat{p})) = \int dq <q|\hat{\rho}\delta(p - \hat{p})|q>, \tag{14}$$

substituting unit operator $\int |n><n|\,dn = 1,$ in (14), and after some calculations one arrives at

$$P_{mom} = \int dq \, dq' \, dp' \, dp'' <q|\hat{\rho}|q'><q'|p'><p'|\delta(p - \hat{p})|p''><p''|q>,$$
$$= \int dq \, dq' \, dp' \, dp'' <q|\hat{\rho}|q'><q'|p'><p'|p''>\delta(p - p'')<p''|q>,$$
$$P_{mom} = \int dq \left\{ \int dq' <q|\hat{\rho}|q'><q'|p><p|q> \right\}. \tag{15}$$

Now DF in phase space will be obtained by comparing Eqs. (13) and (15)

$$P(p,q) = \int dq' <q|\hat{\rho}|q'><q'|p><p|q>,$$
$$= \frac{1}{2\pi\hbar} \int dq' <q|\hat{\rho}|q'> \exp(ip(q'-q)/\hbar). \tag{6}$$

To obtain DF of Eq. (16) one may use $<q|p> = \frac{1}{\sqrt{2\pi\hbar}} \exp(ipq/\hbar)$, but it is not WDF. This is a new DF which by a coordinate transformation

$$q' - q = y \Rightarrow dq' = dy, \tag{17}$$

reduce to Sobouti-Nasiri DF,

$$P_{SN}(p,q) = \frac{1}{2\pi\hbar} \int dy <q|\hat{\rho}|q+y> \exp(ipy/\hbar). \tag{18}$$

Thus, Sobouti-Nasiri DF, which is imaginary, is obtained in this new manner with much less complication than EPS formulation [10]. According to the first assumption, one may apply a new coordinate transformation

$$q' - q \to y,$$
$$\frac{q' + q}{2} \to q, \tag{19}$$

to derive a real DF

$$P_W(p,q) = \frac{1}{2\pi\hbar} \int dy <q - \frac{y}{2}|\hat{\rho}|q + \frac{y}{2}> \exp(\frac{ipy}{\hbar}), \tag{20}$$

which is the same WDF, Eq. (1).

On derivation of Wigner distribution function

## 4. Conclusions

A discussion about derivation of Wigner distribution function is presented in this paper. Extended phase space formulation is considered as a method of derivation of WDF from a unitary transformation of Souboti-Nasiri distribution function. In another word, one may consider a simpler approach for derivation of WDF by assuming the fact that the integrating of phase space distribution functions gives the distribution function in momentum space. Just for WDF, reality of distribution function is considered, too.

**References**

[1] Wigner E P 1932 *Phys. Rev.* **40** 749
[2] Hillery M, O'Connell R F, Scully M O and Wigner E P 1985 *Physics Reports* **106** 121
[3] Kim Y S and Noz M E 1983 *Am. J. Phys.* **51** 368
[4] Kenfack A and Zyczkowski K 2004 *J. Opt. B: Quantum Semiclass. Opt.* **6** 396
[5] Khademi S and Nasiri S 2002 *Iranian Journal of Science & Technology, Transaction A* **20** A1
[6] Mahmoudi M, Salamin Y I and Keitel C H 2005 *Physical Review A* **72** 1
[7] O'Connel R F and Wang L 1985 *Phys. Rev. A* **31** 1707
[8] Glauber R G 1963 *Phys. Rev.* **131** 2766
[9] Kirkwood G 1933 Phys. Rev. **44**, 31
[10] Sobouti Y and Nasiri N 1993 *Int. J. Mod. Phys. B.* **7** 3255
[11] Nasiri S 1995 *Tr. J. Phys.* **18** 1
[12] Khademi S, Nasiri S, Fathi S and Taati F 2006 Study of dissipative system of charged particles by Wigner's functions: *Proc. Quantum Theory and Symmetry IV* (Varna Bulgaria Aug. 2005)
[13] Nasiri S, Khademi S, Bahrami S and Taati F 2006 A Generalized rule for non-commutative operators in extended phase space: *Proc. Quantum Theory and Symmetry IV* (Varna Bulgaria Aug. 2005)